\documentclass[english]{article}
\usepackage[T1]{fontenc}
\usepackage[latin1]{inputenc}
\usepackage{amsmath}
\usepackage{amssymb}

\makeatletter

\newcommand{\noun}[1]{\textsc{#1}}

 \newcommand{\lyxaddress}[1]{
   \par {\raggedright #1 
   \vspace{1.4em}
   \noindent\par}
 }

\usepackage{babel}
\makeatother
\begin{document}

\title{GENERATORS AND ROOTS OF THE QUANTUM LOGIC GATES}

\author{\noun{Rudolf Muradian} and \noun{Diego Frías}}

\maketitle

\lyxaddress{\begin{center}\emph{CEPEDI, Centro de Pesquisas em Informática de
Ilhéus, Departamento de Computação Quântica, Av. Petrobras s/n, Cidade
Nova 45652-570, Ilhéus BA, Brazil.}\end{center}}

\lyxaddress{\begin{center}\emph{UESC, Universidade Estadual de Santa Cruz, Departamento
de Ciências Exatas e Technológicas, Rodovia Ilhéus-Itabuna, km 16,
Salobrinho, 45650-000, Iléus BA, Brazil}\end{center}}

\begin{abstract}
This is an exposition of some new  aspects of quantum logic gates.
At first we established general relations for fundamental quantum
gate $A$ with unique restriction $A^{2}=I$, where $I$ is unit matrix.
The explicit form of the generator and roots of a matrix(gate) $A$
have been found. Then the general results are applied to the particular
cases of one-qubit and multi-qubit quantum gates. Some basic properties
of generators and roots of Pauli and Hadamard gates are demonstrated.


PASC Number(s): 03.-67.-a, 89.70.+c
\end{abstract}

\section{Introduction\label{sec:Introduction}}

The study of mathematical problems of quantum computation and quantum
information theory is an interesting area of research. Qubit (quantum
bit) is the basic unit of quantum information, which is processed
by quantum logic gates. Abstract quantum computer consists of equal
number of $n$ input and $n$ output qubits, connected by unitary
(thus invertible) $2^{n}\times2^{n}$ matrix $U$: $U$$|in>=|out>$
such that $UU^{\dagger}=I=U^{\dagger}U$. Just as classical logical
gates transform classical bits from one state to another, quantum
logic gates change the state of qubits, performing quantum computation.
However, in the quantum case, the unitarity of the logic gates imply
in reversibility $|in>=U^{\dagger}|out>$, an important feature of
quantum computing.

According to the most popular circuit model, first introduced in 1985
by D. Deutsch and succinctly described in a recent publication by
P. Shor \cite{shor} the gate $U$ in turn is made out from elementary
$2\times2$ and $4\times4$ matrices, called quantum gates. In a classical
article of A. Barenco \emph{et al} \cite{barenco} it has been proven
that arbitrary $n$-qubit quantum gate can be implemented by \emph{CNOT}
gates operating on two qubits, and elementary one-qubit gates. Thus,
one-qubit gates combined with \emph{CNOT} form an adequate set of
elementary gates for universal quantum computation. The major advantage
of quantum gates on multiple qubits over usual classical logical gates
is that the number of quantum states which can be processed simultaneously
grows exponentially like $2^{n}$ with the number of qubits considered.
It is important to note the fact that any $n$-qubit gate can be modeled
by one-qubit and two-qubit gates.

In order to construct real physical devices for quantum computation
and quantum communication it is helpful to deal, in some depth, with
mathematical properties of quantum logic gates, the primary tools
for processing the quantum information stored in qubits \cite{shor}-\cite{nielsen}.
In particular, it is interesting to deal with generators and roots
of quantum logic gates. Finding the generator of a logic gate helps
understanding the underlying logical operation, while any root of
a given gate is itself a new gate.

The structure of this article is as follows: In Section \ref{sec:Euler-Formula}
we formulate some general relations for abstract quantum gate $A$
with unique restriction $A^{2}=I$, providing a background for approaching
the characterization of several basic quantum gates in the following
sections. In Section \ref{sec:One-Qubit-Gates} and \ref{sec:Two-Qubit-Gates.}
we apply the general relationships above for the case of one-qubit
gates and to two-qubit gates, respectively. In Section \ref{sec:One-Qubit-Gates}
we also compute the commutators and anti-commutators of the square
roots of elementary quantum gates usually used as building blocks
of more complex quantum operators. The relations of Section \ref{sec:Euler-Formula}
can be applied to any legitimate multi-qubit gate. In order to illustrate
it they are applied to Toffoli and other three-qubit gates \cite{galindo,nielsen,peres}
in Section \ref{sec:Toffoli,Fredkin}. Finally some conclusions are
made in Section \ref{sec:Concluding-Remarks}.

\section{Euler Equation and its Consequences\label{sec:Euler-Formula}}

To calculate generators and roots of different logic gates we need
to build up some general relations, using well known methods of matrix
linear algebra. Let us consider first a self-inverse matrix $A$,
satisfying $A^{2}=I$ or $A=A^{-1}$. The generalized Euler relation
for any such matrix reads as (see, for instance, \cite{nielsen}):

\begin{equation}
e^{\pm i\alpha A}=I\cos\alpha\pm iA\sin\alpha.\label{eq:euler}\end{equation}
Some elegant particular cases of this relation could be noticed:

a) For $\alpha=\pi$ we get

\begin{equation}
e^{\pm i\pi A}=-I.\label{eq:-I}\end{equation}

b) When $\alpha=\pi/2$ we have

\begin{equation}
e^{\pm i\frac{\pi}{2}A}=\pm iA.\label{eq:iA}\end{equation}
 From this relation we obtain

\begin{equation}
A=e^{i\frac{\pi}{2}(I-A)},\label{eq:I-A}\end{equation}
or in a more elegant form 

\begin{equation}
A=i^{I-A}.\label{eq:i I-A}\end{equation}

c) When $\alpha=\pi/4$

\begin{equation}
e^{\pm\frac{\pi}{4}A}=\frac{1}{\sqrt{2}}(I\pm A).\label{eq:I+-A}\end{equation}
From (\ref{eq:i I-A}) it follows that the \emph{n-}th root of arbitrary
self-inverse gate $A$ is provided by the relation:

\begin{equation}
\sqrt[n]{A}=e^{i\frac{\pi}{2n}(I-A)},\,\,\,\, n=1,2,3,...\label{eq:raiz-n}\end{equation}
Using relation \ref{eq:raiz-n} with $n=2$ and keeping in mind equation
\ref{eq:I+-A} the square root of $A$ is straightforwardly computed
as:

\begin{equation}
\sqrt{A}=\frac{1}{\sqrt{2}}(e^{i\pi/4}I+e^{-i\pi/4}A)\label{eq:raiz sq A}\end{equation}
or in equivalent form 

\begin{equation}
\sqrt{A}=\frac{1}{\sqrt{2i}}(iI+A)\label{eq:raiz sq A equivalent}\end{equation}
This general relations will be applied in Section \ref{sec:One-Qubit-Gates}
and Section \ref{sec:Two-Qubit-Gates.} to calculate the square roots
of most commonly used one- and two-qubit quantum gates

The Hermitean matrix $\pi/2(I-A)$ in relation \ref{eq:i I-A} is
the \emph{generator} of the matrix $A$. Such terminology is a reminiscence
of the quantum physics where Hermitean matrix generates evolution
of the unitary matrix. From matrix algebra it is well known that every
unitary matrix $A$ can be represented in exponential form $A=e^{iH}$where
$H$ is a Hermitean matrix called generator of unitary transformation,
since $A^{\dagger}=e^{-iH^{\dagger}}=e^{-iH}$and hence $AA^{\dagger}=1$.
Here $H$ must not be confused with Hadamard gate in the following
sections.

\section{One-Qubit Gates\label{sec:One-Qubit-Gates}}

Now we will consider the application of general expressions \ref{eq:raiz sq A}
and \ref{eq:raiz sq A equivalent} for computation of the square root
of self-inverse matrix $A$ to some particular cases. A quantum gate
is the analogue of a logic gate in a classical circuit model. Typically,
certain primitive gates are used in developing quantum circuits as
elementary building blocks. We are going to consider here the Pauli,
Hadamard and phase gates \cite{galindo},\cite{nielsen}. In particular,
the Pauli gates are of extreme importance in the field of quantum
information theory, because any operator in the one-qubit space can
be written as a linear combination of the following self-inverse Pauli
operators: 

\begin{equation}
I=\left(\begin{array}{cc}
1 & 0\\
0 & 1\end{array}\right),\,\, X=\left(\begin{array}{cc}
0 & 1\\
1 & 0\end{array}\right),\,\, Y=\left(\begin{array}{cc}
0 & -i\\
i & 0\end{array}\right),\,\, Z=\left(\begin{array}{cc}
1 & 0\\
0 & -1\end{array}\right).\label{eq:Pauli}\end{equation}
satisfying 

\begin{equation}
X^{2}=Y^{2}=Z^{2}=I.\label{eq:self-inverse}\end{equation}
These elementary operators act on qubit $|a>,\,\, a\in\{0,1\}$ according
to:

\begin{equation}
\begin{array}{cc}
Identity & I\,|a>=|a>\\
Bit\, flip & X\,|a>=|1\oplus a>\\
Bit-Phase\, flip & Y\,|a>=i(-1)^{a}|1\oplus a>\\
Phase\, flip & Z\,|a>=(-1)^{a}|a>\end{array}\end{equation}
Here $\oplus$is the binary addition operation with properties $0\oplus a=a,$
$1\oplus a=1-a$ and $a\oplus b=a+b-2ab$, where + and - are ordinary
arithmetic operators. 

Another useful gate is the Hadamard gate which makes superposition
of quantum states:

\begin{equation}
H=\frac{1}{\sqrt{2}}(X+Z)=\frac{1}{\sqrt{2}}\left(\begin{array}{cc}
1 & 1\\
1 & -1\end{array}\right)\label{eq:hadamard}\end{equation}
and acts on qubit as follows:

\begin{equation}
H\,|a>=\frac{1}{\sqrt{2}}(X|a>+Z|a>)=\frac{1}{\sqrt{2}}\left(|1\oplus a>+(-1)^{a}|a>\right).\end{equation}
Finally, the phase gates 

\begin{equation}
S=\left(\begin{array}{cc}
1 & 0\\
0 & i\end{array}\right),\,\,\,\,\, T=\left(\begin{array}{cc}
1 & 0\\
0 & e^{i\pi/4}\end{array}\right)\label{eq:phase gate}\end{equation}
multiply basis vectors by a phase, that is:

\begin{equation}
S\,|a>=i^{a}|a>=e^{i\frac{\pi}{2}a}|a>,\,\,\,\,\,\, T\,|a>=e^{i\frac{\pi}{4}a}|a>\end{equation}

\subsection{Roots of One-Qubit Gates\label{sub:Roots-of-One-Qubit}}

Using \ref{eq:raiz sq A} and \ref{eq:raiz sq A equivalent} we can
obtain the square roots of $X,\, Y,\, Z,\, H$ and $S$ gates in the
explicit forms:

a). Pauli matrices:

\begin{equation}
\sqrt{X}=\frac{1}{2}\left(\begin{array}{cc}
1+i & 1-i\\
1-i & 1+i\end{array}\right)=\frac{1}{\sqrt{2i}}(iI+X)=\frac{1}{\sqrt{2i}}\left(\begin{array}{cc}
i & 1\\
1 & i\end{array}\right)\label{eq: sq root X}\end{equation}

\begin{equation}
\sqrt{Y}=\frac{1}{2}\left(\begin{array}{cc}
1+i & -1-i\\
1+i & 1+i\end{array}\right)=\frac{1}{\sqrt{2i}}(iI+Y)=\frac{1}{\sqrt{2i}}\left(\begin{array}{cc}
i & -i\\
i & i\end{array}\right)\label{eq:sq root Y}\end{equation}

\begin{equation}
\sqrt{Z}=\frac{1}{\sqrt{2i}}(iI+Z)=\left(\begin{array}{cc}
1 & 0\\
0 & i\end{array}\right)=S\label{eq:sq root Z}\end{equation}

b). Hadamard gate:

\begin{equation}
\sqrt{H}=\frac{1}{\sqrt{2i}}(iI+H)=\left(\begin{array}{cc}
\cos^{2}\frac{\pi}{8}+i\sin^{2}\frac{\pi}{8} & \frac{1-i}{2\sqrt{2}}\\
\frac{1-i}{2\sqrt{2}} & \sin^{2}\frac{\pi}{8}+i\cos^{2}\frac{\pi}{8}\end{array}\right)\label{eq:sq root H}\end{equation}

c). Phase gate:

\begin{equation}
\sqrt{S}=\frac{1}{\sqrt{2}}(e^{i\pi/4}I+e^{-i\pi/4}S)=\left(\begin{array}{cc}
1 & 0\\
0 & e^{i\pi/4}\end{array}\right)=T\label{eq:sq root Phase}\end{equation}
In the same manner, the corresponding output of the square root gates
on one-qubit state $|a>$ can be written in the compact form:

\begin{equation}
\sqrt{X}\,|a>=\frac{1}{\sqrt{2}}\left(e^{i\frac{\pi}{4}}|a>+e^{-i\frac{\pi}{4}}|1\oplus a>\right)\label{eq:root X}\end{equation}

\begin{equation}
\sqrt{Y}\,|a>=\frac{1}{\sqrt{2}}e^{i\frac{\pi}{4}}\left(|a>+(-1)^{a}|1\oplus a>\right)\label{eq: root Y}\end{equation}
\begin{equation}
\sqrt{Z}\,|a>=\frac{1}{\sqrt{2}}\left(e^{i\frac{\pi}{4}}+e^{-i\frac{\pi}{4}}(-1)^{a}\right)|a>\label{eq:root Z}\end{equation}

\begin{equation}
\sqrt{H}\,|a>=\left(\frac{1}{\sqrt{2}}e^{i\frac{\pi}{4}}+\frac{1}{2}e^{-i\frac{\pi}{4}}(-1)^{a}\right)|a>+\frac{1}{2}e^{-i\frac{\pi}{4}}|1\oplus a>\label{eq:root H}\end{equation}

\subsection{Generators of One-Qubit Operators\label{sub:Generators-of-One-Qubit}}

Pauli matrices in exponential form looks like

\begin{equation}
X=e^{i\frac{\pi}{2}\left(\begin{array}{cc}
1 & -1\\
-1 & 1\end{array}\right)},\,\, Y=e^{i\frac{\pi}{2}\left(\begin{array}{cc}
1 & i\\
-i & 1\end{array}\right)},\,\, Z=e^{i\frac{\pi}{2}\left(\begin{array}{cc}
0 & 0\\
0 & 2\end{array}\right)},\label{eq:Pauli Exp}\end{equation}
This representation allows calculate immediately the square roots
of Pauli matrices which, of course, agree with corresponding root
expressions from \ref{sub:Roots-of-One-Qubit}:

\begin{equation}
\sqrt{X}=e^{i\frac{\pi}{4}\left(\begin{array}{cc}
1 & -1\\
-1 & 1\end{array}\right)}=\frac{1}{\sqrt{2}}\left(\begin{array}{cc}
e^{i\frac{\pi}{4}} & e^{-i\frac{\pi}{4}}\\
e^{-i\frac{\pi}{4}} & e^{i\frac{\pi}{4}}\end{array}\right)=\frac{1}{2}\left(\begin{array}{cc}
1+i & 1-i\\
1-i & 1+i\end{array}\right)\label{eq:sq root X II}\end{equation}

\begin{equation}
\sqrt{Y}=e^{i\frac{\pi}{4}\left(\begin{array}{cc}
1 & i\\
-i & 1\end{array}\right)}=\frac{1}{\sqrt{2}}\left(\begin{array}{cc}
e^{i\frac{\pi}{4}} & -e^{i\frac{\pi}{4}}\\
e^{i\frac{\pi}{4}} & e^{i\frac{\pi}{4}}\end{array}\right)=\frac{1}{2}\left(\begin{array}{cc}
1+i & -1-i\\
1+i & 1+i\end{array}\right)\label{eq:sq root Y II}\end{equation}

\begin{equation}
\sqrt{Z}=e^{i\frac{\pi}{4}\left(\begin{array}{cc}
0 & 0\\
0 & 2\end{array}\right)}=\left(\begin{array}{cc}
1 & 0\\
0 & i\end{array}\right)\label{eq:sq root Z II}\end{equation}
Following the same reasoning the Hadamard matrix can be rewritten
in exponential form as:

\begin{equation}
H=e^{i\frac{\pi}{2}\left(\begin{array}{cc}
1-\frac{1}{\sqrt{2}} & -\frac{1}{\sqrt{2}}\\
-\frac{1}{\sqrt{2}} & 1+\frac{1}{\sqrt{2}}\end{array}\right)}=e^{i\pi\left(\begin{array}{cc}
\sin^{2}\frac{\pi}{8} & -\frac{1}{2\sqrt{2}}\\
-\frac{1}{2\sqrt{2}} & \cos^{2}\frac{\pi}{8}\end{array}\right)}\label{eq:Hadamard Exp}\end{equation}
From this it follows the same expression for the square root of Hadamard
operator as in (\ref{eq:sq root H}):

\begin{equation}
\sqrt{H}=e^{i\frac{\pi}{2}\left(\begin{array}{cc}
\sin^{2}\frac{\pi}{8} & -\frac{1}{2\sqrt{2}}\\
-\frac{1}{2\sqrt{2}} & \cos^{\textsc{\textrm{2}}}\frac{\pi}{8}\end{array}\right)}=\left(\begin{array}{cc}
\mathcal{\cos}^{2}\frac{\pi}{8}+i\sin^{2}\frac{\pi}{8} & \frac{1-i}{2\sqrt{2}}\\
\frac{1-i}{2\sqrt{2}} & \sin^{2}\frac{\pi}{8}+i\cos^{2}\frac{\pi}{8}\end{array}\right)\label{eq:sq root H II}\end{equation}
Let us remind that the eigenvectors of Hadamard matrix for eigenvalues
+1 and -1 are $(\cos\frac{\pi}{8},\sin\frac{\pi}{8})$ and $(-\sin\frac{\pi}{8},\cos\frac{\pi}{8},)$,
respectively (see, for example, equation 112 in \cite{gott}) and
that $\cos\frac{\pi}{8}=\frac{1}{2}\sqrt{2+\sqrt{2}}$ and $\sin\frac{\pi}{8}=\frac{1}{2}\sqrt{2-\sqrt{2}}$.

\subsection{Commutators of the roots of Logic Gates.\label{sub:Commutators-of-the}}

Commutator of two operators $A$ and $B$, also called the \emph{Lie
bracket} is defined by $[A,B]=AB-BA$. The commutators of Pauli matrices
are well known \cite{ballentine} and are given by:

\begin{equation}
[X,Y]=2iZ,\,\,\,[Y,Z]=2iX,\,\,\,[Z,X]=2iY.\end{equation}
We observed that the commutators of the corresponding square root
operators have the same symmetric structure and satisfy the following
relations:

\begin{eqnarray}
[\sqrt{X},\sqrt{Y}]=Z, & \,\,\,[\sqrt{Y},\sqrt{Z}]=X,\,\,\, & [\sqrt{Z},\sqrt{X}]=Y.\end{eqnarray}
Very interesting relations were also obtained for commutators of square
roots of Pauli matrices with Hadamard and square root of Hadamard
operator as shown below:

\begin{eqnarray}
[H,\sqrt{X}]=ie^{-i\pi/4}Y,\,\,\, & [H,\sqrt{Y}]=-e^{i\pi/4}H,\,\,\, & [H,\sqrt{Z}]=-ie^{-i\pi/4}Y\end{eqnarray}

and

\begin{eqnarray}
[\sqrt{H},\sqrt{X}]=\frac{1}{\sqrt{2}}Y,\,\,\, & [\sqrt{H},\sqrt{Y}]=-H,\,\,\, & [\sqrt{H},\sqrt{Z}]=-\frac{1}{\sqrt{2}}Y.\end{eqnarray}
The anticommutator is defined by $\{ A,B\}=AB+BA$. It can be demonstrated
that for square roots of Pauli operators anticommutators are:

\begin{eqnarray}
\{\sqrt{X},\sqrt{Y}\}=Z,\,\,\, & \{\sqrt{Y},\sqrt{Z}\}=X,\,\,\, & \{\sqrt{Z},\sqrt{X}\}=Y.\end{eqnarray}
This could be compared with well known anticommutator relations for
Pauli matrices in Ref. \cite{barenco} and \cite{galindo}.

\section{Two-Qubit Gates.\label{sec:Two-Qubit-Gates.}}

Two-qubit gates lives in the Hilbert space $C^{4}$ and are defined
by tensor product $|ab>=|a>\otimes|b>$, where $a,b\in\{0,1\}$. They
can be processed by two-qubit gates which are the main ingredient
of quantum computer, since calculation eventually is performed through
combination of different qubits. Together with one-qubit gates they
form a universal set sufficient for quantum computing.

The most popular two-qubit gates are \emph{CNOT} and \emph{SWAP} gates.
Since they are self-inverse, $CNOT^{2}=CNOT$ and $SWAP^{2}=SWAP$,
we can apply the general relation \ref{eq:raiz-n} to represent them
in exponential form:

\begin{equation}
CNOT=\left(\begin{array}{cccc}
1 & 0 & 0 & 0\\
0 & 1 & 0 & 0\\
0 & 0 & 0 & 1\\
0 & 0 & 1 & 0\end{array}\right)=e^{i\frac{\pi}{2}\left(\begin{array}{cccc}
0 & 0 & 0 & 0\\
0 & 0 & 0 & 0\\
0 & 0 & 1 & -1\\
0 & 0 & -1 & 1\end{array}\right)}\label{eq: CNOT Exp}\end{equation}

\begin{equation}
SWAP=\left(\begin{array}{cccc}
1 & 0 & 0 & 0\\
0 & 0 & 1 & 0\\
0 & 1 & 0 & 0\\
0 & 0 & 0 & 1\end{array}\right)=e^{i\frac{\pi}{2}\left(\begin{array}{cccc}
0 & 0 & 0 & 0\\
0 & 1 & -1 & 0\\
0 & -1 & 1 & 0\\
0 & 0 & 0 & 0\end{array}\right)}\label{eq: SWAP Exp}\end{equation}
A complete Hermitean $4\times4$ operator basis in this space is provided
by 16 tensor products of Pauli matrices:

\begin{equation}
\begin{array}{cccc}
I\otimes I & I\otimes X & I\otimes Y & I\otimes Z\\
X\otimes I & X\otimes X & X\otimes Y & X\otimes Z\\
Y\otimes I & Y\otimes X & Y\otimes Y & Y\otimes Z\\
Z\otimes I & Z\otimes X & Z\otimes Y & Z\otimes Z\end{array}\label{eq: P P set}\end{equation}
The exponential representation is convenient for calculating roots
of two-qubit gates. For example, the operator $X\otimes X$ has the
following exponential representation:

\begin{equation}
X\otimes X=\left(\begin{array}{cccc}
0 & 0 & 0 & 1\\
0 & 0 & 1 & 0\\
0 & 1 & 0 & 0\\
1 & 0 & 0 & 0\end{array}\right)=e^{i\frac{\pi}{2}\left(\begin{array}{cccc}
1 & 0 & 0 & -1\\
0 & 1 & -1 & 0\\
0 & -1 & 1 & 0\\
-1 & 0 & 0 & 1\end{array}\right)}\end{equation}
and its root is given by:

\begin{equation}
\sqrt{X\otimes X}=e^{i\frac{\pi}{4}\left(\begin{array}{cccc}
1 & 0 & 0 & -1\\
0 & 1 & -1 & 0\\
0 & -1 & 1 & 0\\
-1 & 0 & 0 & 1\end{array}\right)}=\frac{1}{\sqrt{2}}\left(\begin{array}{cccc}
e^{i\pi/4} & 0 & 0 & e^{-i\pi/4}\\
0 & e^{i\pi/4} & e^{-i\pi/4} & 0\\
0 & e^{-i\pi/4} & e^{i\pi/4} & 0\\
e^{-i\pi/4} & 0 & 0 & e^{i\pi/4}\end{array}\right)\end{equation}
All other gates from the set \ref{eq: P P set}can be treated on the
same manner.

\section{Toffoli, Fredkin and Peres Gates.\label{sec:Toffoli,Fredkin}}

A variety of three-qubit gates are known, but we can confine our discussions
to the to the Toffoli, Fredkin \cite{galindo},\cite{nielsen} and
Peres \cite{peres} three-input three-output gates which play an important
role in quantum computation theory.

Our general approach can be as well applied for these three-qubit
gates. From the definition of Toffoli (\emph{CCNOT}) and Fredkin (\emph{F}
or \emph{Controlled-SWAP}) gates 

\begin{equation}
CCNOT\,|a,b,c>=|a,b,c\oplus ab>|\label{eq: Toffoli}\end{equation}

\begin{equation}
F\,|a,b,c>=|a,\overline{b},\overline{c}>\label{eq:Fredkin}\end{equation}
where $\overline{b}=b(1\oplus a)+ca$ and $\overline{c}=c(1\oplus a)+ba$,
and using \ref{eq:raiz sq A} we get the output of the corresponding
root operators:

\begin{equation}
\sqrt{CCNOT}\,|a,b,c>=\frac{1}{\sqrt{2}}\left(e^{i\frac{\pi}{4}}|a,b,c>+e^{-i\frac{\pi}{4}}|a,b,c\oplus ab>\right),\end{equation}

\begin{equation}
\sqrt{F}\,|a,b,c>=\frac{1}{\sqrt{2}}\left(e^{i\frac{\pi}{4}}|a,b,c>+e^{-i\frac{\pi}{4}}|a,\overline{b},\overline{c}>\right).\end{equation}
Peres gate \cite{peres} is defined by transformation:

\begin{equation}
P\,|a,b,c>=|a,b\oplus a,c\oplus ab>\end{equation}
and hence from (\ref{eq:raiz sq A}) it follows that the root of Peres
gate performs the following operation:

\begin{equation}
\sqrt{P}\,|a,b,c>=\frac{1}{\sqrt{2}}\left(e^{i\frac{\pi}{4}}|a,b,c>+e^{-i\frac{\pi}{4}}|a,b\oplus,c\oplus ab>\right).\end{equation}
As an example let us consider the explicit matrix form of the generator
of Toffoli gate. In a three-qubit space the Toffoli gate and his generator
can be represented by the following $8\times8$matrices:

\begin{equation}
CCNOT=\left(\begin{array}{cccccccc}
1 & 0 & 0 & 0 & 0 & 0 & 0 & 0\\
0 & 1 & 0 & 0 & 0 & 0 & 0 & 0\\
0 & 0 & 1 & 0 & 0 & 0 & 0 & 0\\
0 & 0 & 0 & 1 & 0 & 0 & 0 & 0\\
0 & 0 & 0 & 0 & 1 & 0 & 0 & 0\\
0 & 0 & 0 & 0 & 0 & 1 & 0 & 0\\
0 & 0 & 0 & 0 & 0 & 0 & 0 & 1\\
0 & 0 & 0 & 0 & 0 & 0 & 1 & 0\end{array}\right)=e^{i\frac{\pi}{2}\left(\begin{array}{cccccccc}
0 & 0 & 0 & 0 & 0 & 0 & 0 & 0\\
0 & 0 & 0 & 0 & 0 & 0 & 0 & 0\\
0 & 0 & 0 & 0 & 0 & 0 & 0 & 0\\
0 & 0 & 0 & 0 & 0 & 0 & 0 & 0\\
0 & 0 & 0 & 0 & 0 & 0 & 0 & 0\\
0 & 0 & 0 & 0 & 0 & 0 & 0 & 0\\
0 & 0 & 0 & 0 & 0 & 0 & 1 & -1\\
0 & 0 & 0 & 0 & 0 & 0 & -1 & 1\end{array}\right)}\end{equation}
Using this representation it is now easy to calculate the square root
(an other roots) of Toffoli gate. The same can be demonstrated for
Fredkin and Peres gates.

\section{Concluding Remarks\label{sec:Concluding-Remarks}}

We have derived some useful relations for a series of quantum logic
gates which can be assembled into circuits to perform more complicated
quantum operations. For example, \emph{the Controlled-square-root
of-NOT gate} $\sqrt{CNOT}$ \emph{}and square root of \emph{SWAP gate}
$\sqrt{SWAP}$ are frequently used in synthesis of quantum circuits.

In this paper we restricted by consideration only of square roots
of logic gates. But it must be noted that \ref{eq:raiz-n} also enables
to compute higher order (n=3,4,...) roots of corresponding gates.

The hope is that the mathematical results presented here will ade
in the development of future quantum algorithms.

\section*{Acknowledgments}

This work was supported by FAPESB (Research Foundation of the Bahia
State, Brazil).

\end{document}